%% file: ProtonGPDletter.tex
\documentclass[preprint,5p,twocolumn,10pt]{elsarticle}

\usepackage[pdftex,hypertexnames=false,linktocpage=true]{hyperref}
\hypersetup{pdfauthor={Lin Bolang},colorlinks=true,linkcolor=red,anchorcolor=darkgreen,citecolor=green!50!blue,filecolor=darkgreen,urlcolor=green,bookmarksnumbered=true,pdfview=FitB}

\usepackage{lipsum}
\usepackage{mathtools,cuted}

\usepackage{amsmath}

\usepackage{graphicx}
\usepackage{float}
\usepackage{subfigure}

\usepackage{array}
\usepackage{booktabs}
\usepackage{arydshln}
\usepackage{multirow}

\newcommand{\be}{\begin{equation}}  
\newcommand{\ee}{\end{equation}}  

\newcommand{\ba}{\begin{align}}  
\newcommand{\ea}{\end{align}}  

\newcommand{\bp}{b_{\perp}}

\newcommand{\delp}{\Delta_{\perp}}

\newcommand{\ra}{\Big{\rangle}}
\newcommand{\la}{\Big{\langle}}
\newcommand{\ma}{\Big{|}}
\newcommand{\nn}{\nonumber}

\hypersetup{colorlinks=true, citecolor=blue, urlcolor=blue, linkcolor=blue}
\newcommand*{\dif}{\mathop{}\!\mathrm{d}} 
\newcommand*{\iu}{\mathop{}\!\mathrm{i}} 

\input{preamble_math_physics.tex}

\usepackage{braket}

\usepackage{comment}
\begin{document}

\begin{frontmatter}
\title{Generalized parton distributions of gluon in proton: a light-front quantization approach}

\author[imp,ucas,keylab]{Bolang~Lin}%
\ead{linbolang@impcas.ac.cn}

\author[imp,ucas,keylab]{Sreeraj~Nair\corref{cor1}}
\ead{sreeraj@impcas.ac.cn}

\author[imp,ucas,keylab]{Siqi~Xu\corref{cor1}}
\ead{xsq234@impcas.ac.cn}

\author[imp,ucas,keylab]{Zhi~Hu\corref{cor1}}
\ead{huzhi@impcas.ac.cn}

\author[imp,ucas,keylab]{Chandan~Mondal\corref{cor1}}
\ead{mondal@impcas.ac.cn}

\author[imp,ucas,keylab]{Xingbo~Zhao\corref{cor1}}
\ead{xbzhao@impcas.ac.cn}

\author[iowa]{James~P.~Vary}

\author[]{\\\vspace{0.2cm}(BLFQ Collaboration)}

\address[imp]{Institute of Modern Physics, Chinese Academy of Sciences, Lanzhou, Gansu, 730000, China}
\address[ucas]{School of Nuclear Physics, University of Chinese Academy of Sciences, Beijing, 100049, China}
\address[keylab]{CAS Key Laboratory of High Precision Nuclear Spectroscopy, Institute of Modern Physics, Chinese Academy of Sciences, Lanzhou 730000, China}
\address[iowa]{Department of Physics and Astronomy, Iowa State University, Ames, IA 50011, USA}
\cortext[cor1]{Corresponding author}

\begin{abstract}
We solve for the gluon generalized parton distributions (GPDs) inside the proton, focusing specifically on leading twist chiral-even GPDs. We obtain and employ the light-front wavefunctions (LFWFs) of the proton from a light-front quantized Hamiltonian with Quantum Chromodynamics input using basis light-front quantization (BLFQ). Our investigation incorporates the valence Fock sector with three constituent quarks and an additional Fock sector, encompassing three quarks and a dynamical gluon. We examine the GPDs within impact parameter space and evaluate the $x$-dependence of the transverse square radius. We find that the transverse size of the gluon at lower-$x$ is larger than that of the quark, while it exhibits opposite behavior at large-$x$. Using the proton spin sum rule, we also determine the relative contributions of quarks and the gluon to the total angular momentum of the proton. 
\end{abstract}
\end{frontmatter}

\section{Introduction\label{Sec1}}
One of the key challenges in hadron physics is to understand the precise mechanisms by which the non-perturbative structure of the nucleon arises from the theory of quantum chromodynamics (QCD).  A quintessential tool in the investigation of hadron structure is the parton distribution functions (PDFs) which encode information about the longitudinal momentum fraction carried by the active parton. Although PDFs have been utilized widely, a more complete description of the nonperturbative structure of hadrons requires extension of the PDFs into higher dimensional distributions called the generalized parton distributions (GPDs)~\cite{Diehl:2003ny,Ji:1996nm,Radyushkin:1997ki}.  Alongside the transverse momentum dependent parton distributions (TMDs), the GPDs have been established as an integral component in nucleon tomography~\cite{Lin:2020rxa} by unveiling the 3-dimensional structure of the nucleon.  In addition to the longitudinal momentum fraction ($x$), the GPDs also depend on the  square of the total momentum transferred ($t$) and the longitudinal momentum transferred ($\zeta$),  also  called the skewness variable. In the forward limit $t=0$ and $\zeta=0$, the GPDs reduce to the one dimensional ordinary PDFs. The moments (integration over $x$) of GPDs correspond to nucleon form factors. When the skewness $\zeta$ is zero, the Fourier transform of the GPDs with respect to the transverse momentum transfer $\Delta_{\perp}$ yields the impact parameter dependent parton distributions (ipdpdfs)~\cite{Burkardt:2000za,Burkardt:2002hr}. These distributions show how partons of a particular longitudinal momentum are distributed in the transverse position, also known as the impact parameter $b_{\perp}$ space. The ipdpdfs adhere to certain positivity restrictions and, unlike the GPDs, have a probabilistic interpretation~\cite{Ralston:2001xs}.

Experimentally, GPDs are accessible via the hard exclusive reactions, such as the deeply virtual Compton scattering (DVCS)~\cite{Ji:1996nm,Radyushkin:1997ki,Diehl:2015uka,Radyushkin:1996nd}, deeply virtual meson production (DVMP)~\cite{Radyushkin:1996ru,Collins:1996fb}, wide-angle Compton scattering (WACS)~\cite{Radyushkin:1998rt,Diehl:1998kh}, and also single diffractive hard exclusive processes (SDHEPs)~\cite{Qiu:2022pla}.  Jefferson Lab (JLab) has produced a significant amount of exclusive measurements through the use of extensive data sets~\cite{CLAS:2018ddh,CLAS:2021gwi,Georges:2017xjy,JeffersonLabHallA:2022pnx}. The forthcoming Electron-Ion Collider (EIC)~\cite{Accardi:2012qut,AbdulKhalek:2021gbh} at Brookhaven National Lab (BNL) and Electron-Ion Collider in China (EIcC)~\cite{Anderle:2021wcy} are projected to generate extensive additional data. Besides experiments, theoretical studies have also made significant progress. Nevertheless, the nonperturbative properties of GPDs prohibit their direct computation from the first principles of QCD at the present time. While there have been some calculations of GPDs based on Euclidean-space lattice results, the methodology is still nascent in its development~\cite{Ji:2013dva,Ji:2020ect,Alexandrou:2020zbe,Lin:2021brq}.  Various nonperturbative methods have been utilized to examine the properties of GPDs from a more phenomenological perspective. These methods include the MIT bag model~\cite{Ji:1997gm}, the chiral quark-soliton model~\cite{Goeke:2001tz,Ossmann:2004bp}, the light-front constituent quark model~\cite{Boffi:2002yy,Scopetta:2003et,Choi:2001fc,Choi:2002ic}, NJL model~\cite{Mineo:2005vs}, the color glass condensate model~\cite{Goeke:2008jz}, the Bethe-Salpeter approach~\cite{Tiburzi:2001je,Theussl:2002xp} and the meson cloud model~\cite{Pasquini:2006dv,Pasquini:2006ib}.

In comparison to quark distributions, the study of gluon GPDs is relatively limited. The gluon distributions have an impact on the calculated cross-section of a process dominated by the gluon-initiated channel. On the other hand, gluons have a considerable influence on the mass decomposition of the proton~\cite{Hatta:2018sqd,Tanaka:2018nae,Ji:1994av,Lorce:2017xzd,Ji:1995sv,Burkert:2023wzr,Rodini:2020pis,Metz:2020vxd}. In the study of deep inelastic scattering (DIS) processes, the gluon distributions and fragmentation functions also encode essential information about the proton~\cite{Xie:2022lra}. The leading twist gluon GPDs in a quark model  within perturbative QCD were shown in Ref.~\cite{Meissner:2007rx}. A parametrization of the chiral-even GPDs for gluons with non-zero skewness in a perturbative QCD framework were shown in Ref.~\cite{Kriesten:2021sqc}. With the light-cone spectator model, the leading twist gluon GPDs and the gluon angular momentum inside the proton were studied in Ref.~\cite{Tan:2023kbl}. The lowest two Mellin moments of the gluon GPDs and the gluonic contribution to the nucleon spin have been explored within a model based on the Rainbow-Ladder truncation of the Dyson-Schwinger equations (DSE) of QCD~\cite{Tandy:2023zio} as well as using lattice QCD at physical pion mass~\cite{Alexandrou:2020sml}. 

In this work, we calculate the leading twist chiral-even gluon GPDs within basis light-front quantization (BLFQ). BLFQ is a nonperturbative framework that is proving to be effective in solving problems related to relativistic many-body bound states in quantum field theories~\cite{Vary:2009gt,Zhao:2014xaa,Nair:2022evk,Wiecki:2014ola,Li:2015zda,Jia:2018ary,Tang:2019gvn,Lan:2019vui,Mondal:2019jdg,Xu:2021wwj,Lan:2021wok,Kuang:2022vdy}. In the BLFQ framework we ultilize an effective light-front Hamiltonian and solve for its mass eigenstates~\cite{Vary:2009gt}. We take into account the baryon Fock sector, which includes one gluon ($\ket{qqqg}$), in addition to the valence Fock sector consisting of three quarks ($\ket{qqq}$). We consider the QCD light-front interaction~\cite{Brodsky:1997de} that applies to both of these Fock sectors, along with the model confining potentials that act in both the transverse and longitudinal directions~\cite{Li:2015zda}.  We calculate the quark and gluon contribution to the total angular momentum ($J_{q/g}$) of the proton by using the spin sum rule that connects the moments of the GPDs to $J_{q/g}$~\cite{Ji:1996ek}. We also investigate the gluon GPDs in the impact parameter space~\cite{Burkardt:2000za,Burkardt:2002hr} by Fourier transforming from the transverse $\Delta_{\perp}$  to $b_{\perp}$ space.

\section{Proton LFWFs in the BLFQ framework\label{Sec2}}
In light-front quantum field theory, bound states are obtained by solving the mass eigenvalue equation
\begin{equation}
	\left(P^+ P^- - P_\perp^{~2}\right)\ket{\Psi} = M^2 \ket{\Psi},
	\label{mass eigenvalue equation}
\end{equation} 
where $P^-$, $P^+$, $P_\perp$ and $M$ represent the light-front Hamiltonian, longitudinal momentum, transverse momentum and invariant mass, respectively. At constant light-front time $x^+ \equiv x^0+x^3$, a baryon state is expressed using various Fock sectors:
\begin{equation}
	\ket{\Psi}=\psi_{qqq}\ket{qqq}+\psi_{qqqg}\ket{qqqg}+\cdots
	\label{Fock}
\end{equation}
$\psi_{\cdots}$ denotes the light-front amplitudes associated with the Fock component $\ket{\cdots}$. Numerical calculations require truncating Fock sector expansions to a countable Fock space via Eq.~\eqref{Fock}, in this case including one dynamical gluon. Thus, at the model scale, the proton is described by light-front amplitudes of valence quarks $\psi_{uud}$ and three quarks with one dynamical gluon $\psi_{uudg}$.

We use a light-front Hamiltonian, $P^-=P^-_0+P^-_{I}$, in which $P^-_0$ refers to the light-front QCD Hamiltonian associated with the $\ket{qqq}$ and $\ket{qqqg}$ Fock states of the proton, while $P^-_I$ denotes a model Hamiltonian for the confining interaction potential. We truncate  the light-front QCD Hamiltonian to a single dynamical gluon Fock sector by employing the gauge $A^+=0$~\cite{Lan:2021wok,Brodsky:1997de,Xu:2022abw}:
\begin{equation}
	\begin{split}
		P_0^-= &\int \mathrm{d}x^- \mathrm{d}^2 x^{\perp} \Big\{\frac{1}{2}\bar{\psi}\gamma^+\frac{m_{0}^2+(i\partial^\perp)^2}{i\partial^+}\psi \\
		&+\frac{1}{2}A_a^i\left[m_g^2+(i\partial^\perp)^2\right] A^i_a +g_c\bar{\psi}\gamma_{\mu}T^aA_a^{\mu}\psi \\
		&+ \frac{1}{2}g_c^2\bar{\psi}\gamma^+T^a\psi\frac{1}{(i\partial^+)^2}\bar{\psi}\gamma^+T^a\psi \Big\}.
	\end{split}
\label{hami}
\end{equation}
Here, $\psi$ and $A^\mu$ correspond to the quark and gluon fields, respectively, and $T^a$ represents one half times the Gell-Mann matrix, expressed as $T^a=\lambda^a/2$. We denote the bare quark mass by $m_0$ and the model gluon mass by $m_g$. In Eq.~\eqref{hami}, the initial two terms represent the kinetic energies of the quark and gluon, respectively. The final two terms in Eq.~\eqref{hami} correspond to the vertex interaction and the instantaneous interaction, both of which involve the coupling constant $g_c$. Although the gluon mass is identically zero in QCD, an effective gluon mass, motivated in part by the Renormalization Group Project for Effective Particles~\cite{Glazek:2017rwe}, is employed in our model to reproduce the nucleon mass and form factors (FFs)~\cite{Xu:2022abw}. We incorporate a mass counter term, $\delta m_q=m_0-m_q$, for quarks in the leading Fock sector to accommodate quark mass corrections arising from fluctuations into higher Fock sectors, where $m_q$ represents the renormalized quark mass. In the case of the vertex interaction, we introduce an independent quark mass, $m_f$, in accordance with Ref.~\cite{Glazek:1992aq,Burkardt:1998dd}.

The confining interaction potential within the leading Fock sector, encompassing both transverse and longitudinal confining potentials, can be expressed as follows~\cite{Lan:2021wok,Li:2015zda}:
\begin{equation}
	\begin{split}
		\label{eqn:Hc}
		P_{I}^-P^+=\frac{\kappa^4}{2}\sum_{i\neq j}^{3}\left\{{{r}_{ij\perp}}^{2}-\frac{\partial_{x_i}\left(x_i{x_j}\partial_{x_j}\right)}{\left( m_i+m_j\right)^2}\right\}.
	\end{split}
\end{equation}
Here, $\kappa$ represents the confinement strength, and $r_{ij\perp}=\sqrt{x_i x_j}(r_{i\perp}-r_{j\perp}^{~\prime})$ denotes the intrinsic holographic variable between the struck parton and the spectator~\cite{Brodsky:2014yha}.

In the BLFQ approach~\cite{Vary:2009gt}, we employ the two dimensional harmonic oscillator (2D-HO) basis functions with scale parameter $b$ in $\Phi_{nm}({p}_\perp;b)$ to describe transverse degrees of freedom~\cite{Li:2015zda}. The 2D-HO basis function is defined by radial ($n$) and angular ($m$) quantum numbers. For the longitudinal motion of Fock partons, we utilize plane-wave basis functions. The longitudinal motion is confined to a box of length $2L$ with anti-periodic (periodic) boundary conditions for fermions (bosons). Longitudinal momentum is expressed as $p^+=\frac{\pi}{L}k$, with $k$ being a half-integer (integer) for fermions (bosons) representing the longitudinal momentum. We neglect the zero mode for the bosons. Each single-particle basis state then has four quantum numbers, $\ket{\alpha_i}=\ket{k_i,n_i,m_i,\lambda_i}$, with $\lambda$ indicating the dimensionless discrete value of the light-front helicity of the parton. The multi-body basis states are defined as the direct product of single-particle basis states, $\ket{\alpha}=\otimes_i \ket{\alpha_i}$. 

We enforce an overall color singlet structure to our many-body basis space. In instances where Fock sectors permit the existence of multiple color-singlet states, it is necessary to introduce an auxiliary label to uniquely identify each color singlet state. The leading Fock sector $\ket{qqq}$ has only one color singlet state, where all the three quarks have different colors. Even though, the identical flavored quarks have the same quantum numbers such as spin, momentum, etc., they are distinguishable due to different color quantum numbers in the $\ket{qqq}$ Fock sector. Thus, antisymmetrization of identical quarks is not needed within the $\ket{qqq}$ Fock sector. On the other hand, the $\ket{qqqg}$ configuration possesses two distinct color-singlet states and the identical flavored quarks can have the same quantum numbers including color. However, within our current basis truncation, a very small portion of basis space having two identical quarks might reasonably be assumed to have a minimal effect on the proton observables. We therefore believe it is a reasonable approximation to neglect antisymmetrization of identical quarks within the $\ket{qqqg}$ Fock sector.

Our many-body basis states possess well-defined total angular momentum projection:
\begin{equation}
	M_J=\sum_i^N (m_i+\lambda_i),
	\label{total angular momentum projection}
\end{equation}
where $N$ represents the number of particles in the Fock sector. We introduce two truncation parameters, $\mathcal{K}$ and $\mathcal{N}$, for the longitudinal and transverse directions, respectively. Along with our Fock space truncation, these two parameters and $b$ fix our model scale. The dimensionless variable $\mathcal{K}\equiv \sum_i k_i$ characterizes the bound state's longitudinal momentum $P^+$. The longitudinal momentum fraction $x$ is defined as $x_i \equiv p_i^+/P^+=k_i/\mathcal{K}$ for the $i^\mathrm{th}$ parton. $\mathcal{K}$ represents the resolution of our method in the longitudinal direction and impacts the resolution of the PDFs. The $\mathcal{N}$ truncation, defined by $\sum_i(2n_i+|m_i|+1)\le \mathcal{N}$, allows for the factorization of transverse center-of-mass motion~\cite{Wiecki:2014ola,Zhao:2014xaa}. This truncation is associated with infrared (IR) and ultraviolet (UV) cutoffs in the transverse direction. In momentum space, the IR cutoff $\Lambda_{\rm IR} \propto b/\sqrt{\mathcal{N}}$, and the UV cutoff $\Lambda_{\rm UV} \propto b\sqrt{\mathcal{N}}$~\cite{Zhao:2014xaa}.

Upon diagonalizing the light-front Hamiltonian matrix within our framework, we obtain the mass spectra $M^2$ (eigenvalues) and the corresponding LFWFs in the momentum space (eigenstates) as:
\begin{equation}
	\Psi^{M_J}_{N,\{\lambda_i\}}(\{x_i,{p_i}_{\perp}\})=\sum_{\{n_i m_i\}}\psi^{M_J}_{N}(\{{\alpha}_i\})\prod_{i=1}^{N}  \Phi_{n_i m_i}({p_i}_{\perp},b).
	\label{light-front wave function}
\end{equation}
Here, $\psi^{M_J}_{N=3}({\alpha_i})$ and $\psi^{M_J}_{N=4}({\alpha_i})$ are the components of the eigenvectors associated with the Fock sectors $\ket{uud}$ and $\ket{uudg}$, respectively.

\section{Generalized parton distributions\label{Sec3}}
There are four chiral-even (helicity-conserving) gluon GPDs at leading twist, denoted as $H_g$, $E_g$, $\widetilde{H}_g$, and $\widetilde{E}_g$. These GPDs are defined by considering off-forward matrix elements of light-front bilocal currents~\cite{Ji:1998pc}. In the light-front gauge, where $A^+ = 0$, the gauge link is unity, enabling the definition of the leading twist gluon GPDs as follows~\cite{Diehl:2003ny}:
\begin{align}
F^{g}_{\Lambda,\Lambda'} &= \frac{1}{\bar{P}^+}\int \frac{dz^-}{2\pi} e^{ixP^+z^-} \nn \\
 &\times \la P',\Lambda' \ma G^{+\mu} \left(-\frac{z}{2} \right) G_{\mu}^{+}  \left(\frac{z}{2} \right)  \ma P,\Lambda \ra \Big{|}_{\substack{z^+=0\\z^{\perp}=0}} \nn \\
&=  \frac{1}{2\bar{P}^+} \bar{u}(P',\Lambda')\left[ H^g\left( x,\zeta,t\right) \gamma^+ \right. \nn \\
&+\left. E^g \left( x,\zeta,t\right) \frac{i\sigma^{+\alpha} \left(-\Delta_{\alpha}\right)}{2M}\right] u(P,\Lambda),  \\
\widetilde{F}^{g}_{\Lambda,\Lambda'} &= \frac{-i}{\bar{P}^+} \int \frac{dz^-}{2\pi} e^{ixP^+z^-} \nn \\
&\times \la P',\Lambda' \ma G^{+\mu} \left(-\frac{z}{2} \right) \widetilde{G}_{\mu}^{+}  \left(\frac{z}{2} \right)  \ma P,\Lambda \ra \Big{|}_{\substack{z^+=0\\z^{\perp}=0}} \nn \\
&=  \frac{1}{2\bar{P}^+}\bar{u}(P',\Lambda')\left[ \widetilde{H}^g\left( x,\zeta,t\right) \gamma^+ \gamma_5 \right. \nn \\
&+ \left. \widetilde{E}^g \left( x,\zeta,t\right) \frac{\gamma_5 \left(-\Delta_{\alpha}\right)}{2M}\right] u(P,\Lambda),
\end{align}
where, $G^{+\mu}(x) = \partial^{+}A^{\mu}(x)$ represents the gluon field tensor in the light-cone gauge, while the dual field strength is given by $\widetilde{G}^{\alpha \beta}(x) = \frac{1}{2}\epsilon^{\alpha \beta \gamma \delta }G_{\gamma \delta}(x)$. The momenta of the proton initial and final states are denoted by $P$ and $P'$, respectively, with their corresponding helicities represented as $\Lambda$ and $\Lambda'$. The light-front spinor for the proton is  denoted by $u(P,\Lambda)$. The momentum transfer in this scenario is expressed as $\Delta^{\mu} = P^{'\mu}-P^{\mu}$, while the skewness is given by $\zeta = -\frac{\Delta^+}{2\bar{P}^+}$. We use a symmetric frame wherein $P^{'\perp} = \Delta^{\perp}/2$ and $P^{\perp} = -\Delta^{\perp}/2$. The average momentum $\bar{P}^\mu=(P^\mu + P^{'\mu})/2$. In this work, we consider the zero skewness case ($\zeta =0$) and thus only study the kinematic region corresponding to $\zeta < x < 1 $ also called the DGLAP region. The invariant momentum transfer in the process is denoted by $t=\Delta^2$ and for zero skewness, we have $t=-\Delta_\perp^2$. Note that one has to consider nonzero skewness to compute $\widetilde{E}^g$. The remaining three chiral-even gluon GPDs at leading twist  can be represented in terms of diagonal ($N \rightarrow N$) overlap of LFWFs as follows:
\begin{align}
H^g\left( x,0,t\right) &= \sum_{ \{\lambda_i \}} \int\left[\dif\mathcal{X}\dif\mathcal{P}_\perp\right] \delta(x-x_1)\nn \\
\times &  ~\Psi_{ 4,\{\lambda_i \}}^{\uparrow*} (\{x'_i,{p'}_{i\perp}\})
\Psi_{ 4,\{\lambda_i \}}^{\uparrow} (\{x_i,{p}_{i\perp}\}) ,\label{gpd_eq1}  
\end{align}
\begin{align}
E^g\left( x,0,t\right) &= \frac{2M}{\Delta_1-i\Delta_2}\sum_{ \{\lambda_i \}} \int\left[\dif\mathcal{X}\dif\mathcal{P}_\perp\right] \delta(x-x_1)\nn \\
\times &  ~\Psi_{ 4,\{\lambda_i \}}^{\uparrow*} (\{x'_i,{p'}_{i\perp}\})
\Psi_{ 4,\{\lambda_i \}}^{\downarrow} (\{x_i,{p}_{i\perp}\}) ,\label{gpd_eq2}
\end{align}
\begin{align}
\widetilde{H}^g\left( x,0,t\right) &= \sum_{ \{\lambda_i \}} \int\left[\dif\mathcal{X}\dif\mathcal{P}_\perp\right] \delta(x-x_1)\nn \\
\times &  ~\lambda_4 \Psi_{ 4,\{\lambda_i \}}^{\uparrow*} (\{x'_i,{p'}_{i\perp}\})
\Psi_{ 4,\{\lambda_i \}}^{\uparrow} (\{x_i,{p}_{i\perp}\}) ,\label{gpd_eq3}
\end{align}
where the longitudinal and transverse moemnta of the struck parton are $x'_1=x_1$ and ${p'}_{1\perp} ={p}_{1\perp} + (1-x_1){\Delta}_{\perp}$ respectively. The $M$ is the proton mass in Eq.~\eqref{gpd_eq2}. The spectator momenta are $x'_i=x_i$ and ${p'}_{1\perp} ={p}_{1\perp} + x_1{\Delta}_{\perp}$. The shorthand notation used for the  integration measure is as follows:
\begin{equation}
	\begin{split}
		\left[\dif\mathcal{X}\dif\mathcal{P}_\perp\right]&\equiv \prod_{i=1}^N\left[\frac{\dif x_i \dif^2k_{i\perp}}{16\pi^3}\right]16\pi^3\delta\left(1-\sum_{i=1}^N x_i\right)\\
		&\times \delta^2\left(\sum_{i=1}^Nk_{i\perp}\right).
	\end{split}
\end{equation}

We study the gluon GPDs in the impact parameter space by performing a two dimensional Fourier transform (FT) with respect to the transverse momentum transfer~\cite{Burkardt:2002hr,Diehl:2002he} at zero skewness:
\begin{equation}
	\mathcal{F}\left(x,b_{\perp}\right)=\int \frac{\dif^2\Delta_\perp}{\left(2\pi\right)^2}e^{-\iu \delp \bp}F\left(x,\zeta=0,t=-\delp^2\right),
\end{equation}
where $F=( H^g,E^g,\widetilde{H}^g)$. The impact parameter variable $\bp$ in the transverse coordinate space is conjugate to the transverse momentum transfer ${\Delta}_{\perp}$. $\bp$ denotes the transverse separation between the struck parton and the nucleon's transverse center of momentum, expressed as $\sum_i x_i b_{\perp i}$~\cite{Diehl:2003ny} with summation over parton indices. The relative distance between the struck parton and the spectator partons' center of momentum is $\bp/(1-x)$, enabling the estimation of the bound state's transverse size~\cite{Diehl:2004cx}. The Fourier transform of  GPD $H$, represented by the function $\mathcal{H}$, is of particular interest as it characterizes the parton number density with longitudinal momentum fraction $x$ at a specific transverse distance $\bp$ inside the nucleon~\cite{Burkardt:2000za}. Thus, we can define the transverse parton density's $x$-dependent squared radius~\cite{Lorce:2007fa}
\begin{equation}
	\begin{split}
		\left\langle b_\perp^2 \right\rangle^{q/g} \left(x\right) = \frac{\int \dif^2 {b}_\perp \left({b}_\perp\right)^2\mathcal{H}\left(x,0,{b}_\perp\right)}{\int \dif^2 {b}_\perp \mathcal{H}\left(x,0,{b}_\perp\right)},
		\label{theoretical parton squared radius}
	\end{split}
\end{equation}
which can also be written through the GPD $H^{q/g}\left(x,0,t\right)$ as:
\begin{equation}
	\begin{split}
		\braket{b_\perp^2}^{q/g}(x)=-4\frac{\partial}{\partial (-t)}\mathrm{ln}H\left(x,0,t\right).
	\end{split}
\end{equation}

Meanwhile the $\mathcal{E}(x,b_\perp)$ illustrates a deformation of the density of the unpolarized parton in the transversely polarized proton~\cite{Burkardt:2002hr}. $\widetilde{\mathcal{H}}$ is responsible for the density of longitudinally polarized parton in the unpolarized proton.

We also determine the contribution of each parton flavor to the total spin of the proton by utilizing the spin sum rule. The nucleon spin sum rule relates the first moment of GPDs to the total angular momentum of the proton~\cite{Ji:1996ek}
\begin{equation}
	\begin{split}
		J^z=\frac{1}{2}\int \dif x ~x \left[H(x,0,0)+E(x,0,0)\right].
	\end{split}
	\label{jirule}
\end{equation}

\begin{figure*}[hbt!]
	\begin{center}
		\includegraphics[width=0.325\linewidth]{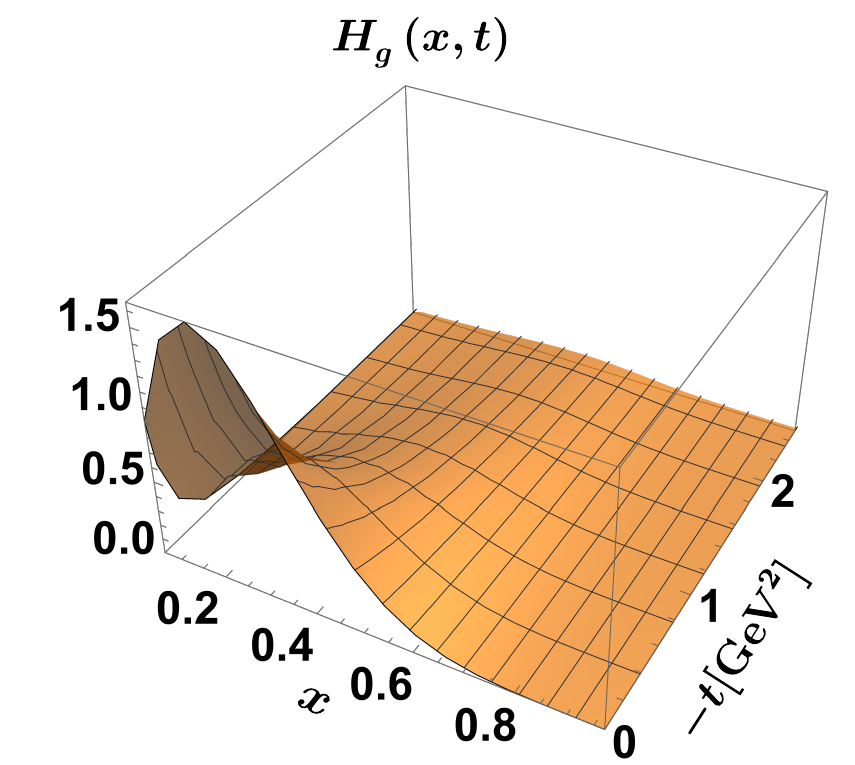}
		\includegraphics[width=0.325\linewidth]{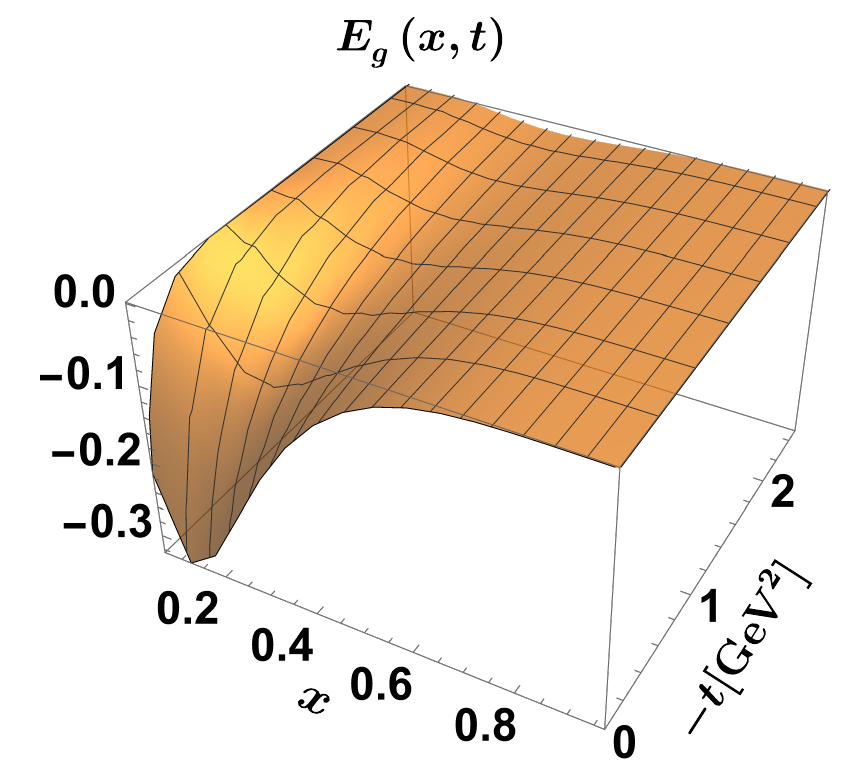}
		\includegraphics[width=0.325\linewidth]{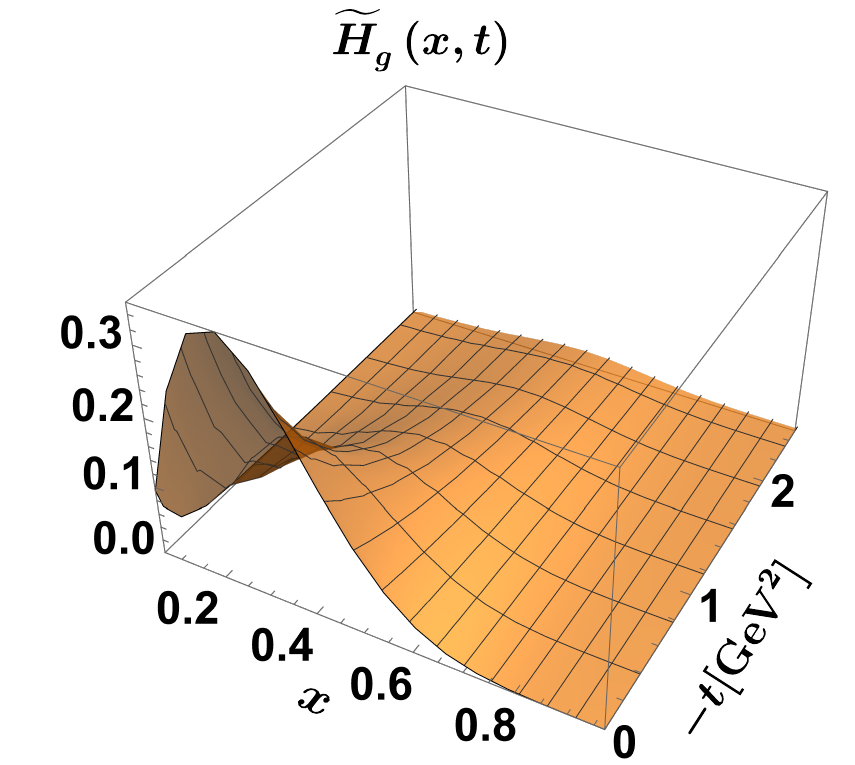}
		\caption{3D plots for the three non-zero chiral-even gluon GPDs, $H_g(x,t)$, $E_g(x,t)$, and $\widetilde{H}_g(x,t)$ as functions of $x$ and $-t$ at the scale $\mu_0^2=0.23 \sim 0.25~\mathrm{GeV}^2$. All results are shown at zero skewness.}
		\label{gluon GPD in momentum space}
	\end{center}
\end{figure*}
\begin{figure}[!ht]
	\begin{center}
		\includegraphics[width=0.98\linewidth]{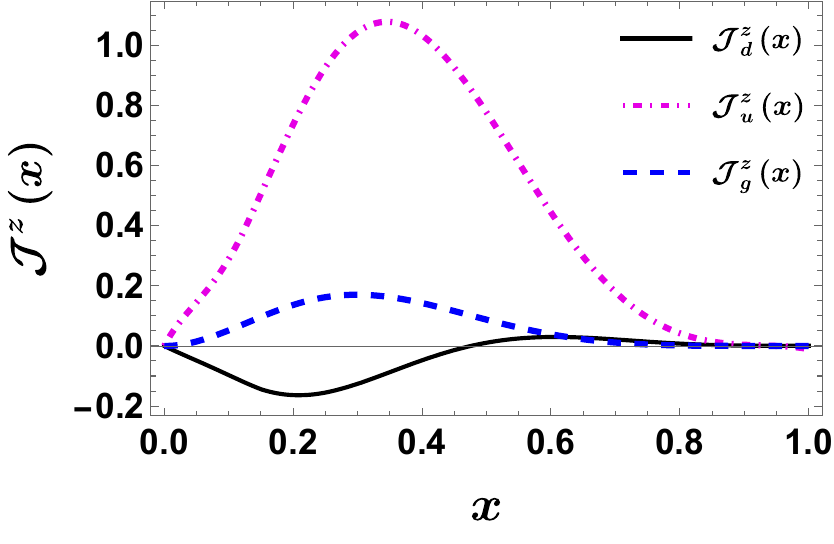}
		\caption{Plot of the density function $\mathcal{J}^z(x)$ vs $x$ for valence quarks and dynamical gluon. The black solid curve is for the $d$-quark, the dot-dashed magenta curve is for the $u$-quark and the blue dashed curve is for the gluon.}
		\label{density Jz in valence quarks and dynamical gluon}
	\end{center}
\end{figure}

\section{Numerical results\label{Sec4}}
The transverse and longitudinal truncation parameters are set to $\mathcal{N}=9$ and $\mathcal{K} =16.5$, respectively, throughout the entire calculation. We choose the harmonic oscillator scale parameter $b=0.70~\mathrm{GeV}$ and the UV cutoff for the instantaneous interaction $b_{\mathrm{inst}}=3.00~\mathrm{GeV}$. The model parameters of the Hamiltonian are $\{m_u, m_d, m_g, \kappa, m_f, g_c\} = \{0.31, 0.25, 0.50, 0.54, 1.80, 2.40\}$, with all values in GeV unit, except for $g_c$. These values are obtained by fitting the proton mass ($M$), its electromagnetic properties, and flavor FFs~\cite{Xu:2022abw}.

With the above parameters set, we numerically solve the eigenvalue equation to obtain the corresponding boost-invariant LFWF, denoted by $\Psi^{M_J}_{N,{\lambda_i}}(\{x_i,p_i^{\perp}\})$ at the model scale $\mu_0^2=0.23 \sim 0.25~\mathrm{GeV}^2$~\cite{Xu:2022abw}. It is important to note that LFWFs should exhibit parity symmetry (P); however, this is disrupted by Fock space truncation. Nevertheless, mirror parity, represented by $\hat{P}_x =\hat{R}_x(\pi)P$~\cite{Brodsky:2006ez}, can be used as a substitute for parity. When applying the mirror parity transformation, eigenvectors corresponding to $M_J=-\frac{1}{2}$ can be derived from the eigenvector associated with $M_J=\frac{1}{2}$ using the following relationship: $\psi^{\downarrow}_{N}\left({\{x_i,n_i,m_i,\lambda_i\}}\right)= (-1)^{\sum_i m_i+c(N)} \psi^{\uparrow}_{N}\left({\{x_i,n_i,-m_i,-\lambda_i\}}\right)$, where $c(3)=1$ and $c(4)=0$. We subsequently utilize the obtained LFWFs in conjunction with the expressions from Eqs.~\eqref{gpd_eq1} to \eqref{gpd_eq3} to calculate the gluon GPDs in both momentum space and impact parameter space.

\subsection{The gluon GPDs in the momentum space\label{Sec4.1}}
In this study, we focus on the case of zero skewness and consequently omit the skewness argument when writing the GPDs. We present three-dimensional (3D) plots for the three non-zero chiral-even gluon GPDs, $H_g(x,t)$, $E_g(x,t)$, and $\widetilde{H}_g(x,t)$, in Fig.~\ref{gluon GPD in momentum space}. Both $H_g(x,t)$ and $\widetilde{H}_g(x,t)$ exhibit positive peaks along the $-t$ direction, while $E_g(x,t)$ displays negative peaks. The maximum peak value for all three distributions occurs at the forward limit $-t=0.00~\mathrm{GeV}^2$. Notably, the largest peak for $H_g(x,t)$ is significantly greater than those for $E_g(x,t)$ and $\widetilde{H}_g(x,t)$. The maximum peak magnitudes for both $E_g(x,t)$ and $\widetilde{H}_g(x,t)$ are comparable. We also notice that the magnitudes of all the GPDs decrease and the peaks along $x$ move towards larger values of $x$ as the momentum transfer $-t$ increases similar to that observed for the quark GPDs in the proton~\cite{Ji:1997gm,Scopetta:2002xq,Petrov:1998kf,Penttinen:1999th,Boffi:2002yy,Boffi:2003yj,Vega:2010ns,Chakrabarti:2013gra,Mondal:2015uha,Chakrabarti:2015ama,Mondal:2017wbf,deTeramond:2018ecg,Xu:2021wwj} as well as in the light mesons~\cite{Kaur:2018ewq,deTeramond:2018ecg,Zhang:2021mtn,Kaur:2020vkq}. The forward limits of $H_g(x,0)$ and $\widetilde{H}_g(x,0)$ correspond to ordinary spin-independent and spin-dependent gluon PDFs, respectively, which have been reported within our BLFQ approach in Ref.~\cite{Xu:2022abw}. We note that $E_g(x,t)$ decouples in the forward limit, as demonstrated in Eq.~\eqref{gpd_eq2}, so no such limit exists for $E_g(x,t)$~\cite{Diehl:2003ny}.

When comparing with the results from the light-cone spectator model~\cite{Tan:2023kbl}, a significant difference can be observed. Contrary to our case, where both $H_g(x,t)$ and $\widetilde{H}_g(x,t)$ represent positive definite quantities, the spectator model displays regions around small $x$ that are negative. However, the qualitative nature of $E_g(x,t)$ as shown in Ref.~\cite{Tan:2023kbl} aligns with our findings.

\begin{figure*}[htp]
	\begin{center}
		\includegraphics[width=0.325\linewidth]{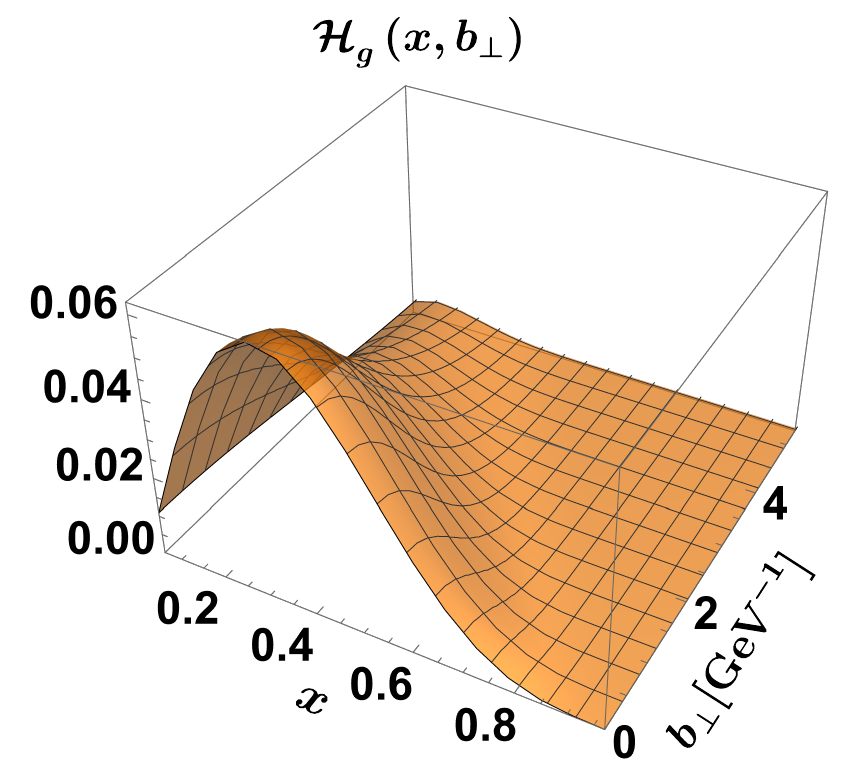}
		\includegraphics[width=0.325\linewidth]{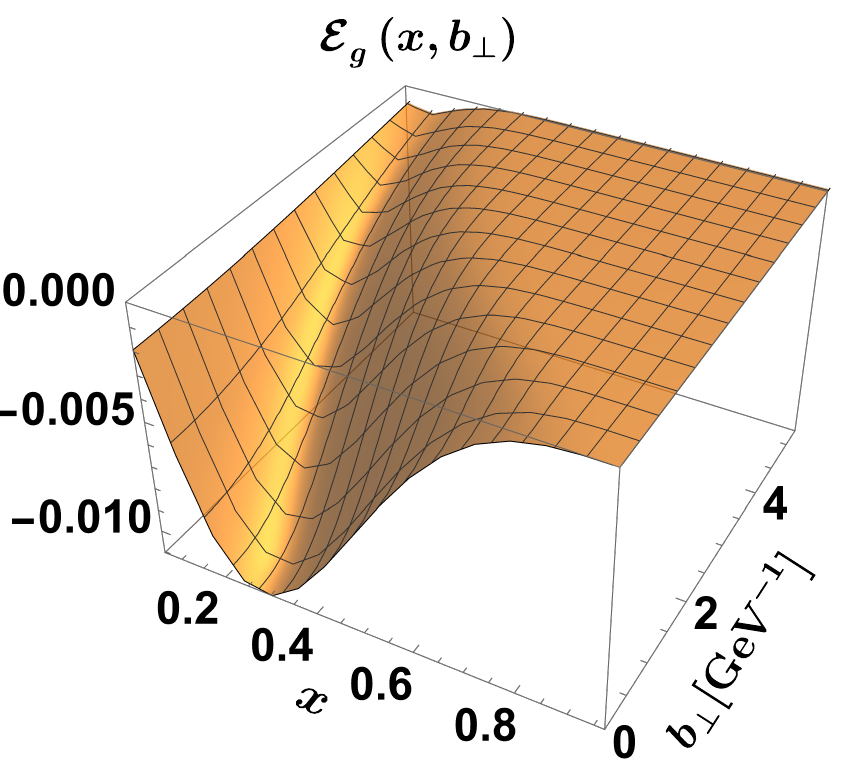}
		\includegraphics[width=0.325\linewidth]{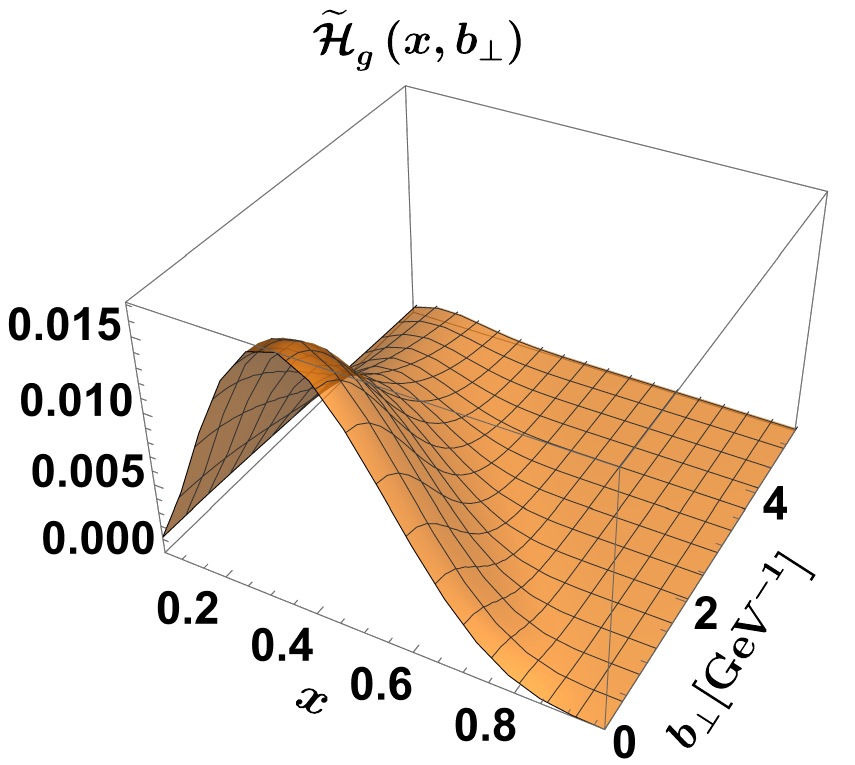}
		\caption{3D plots for the FT of the gluon GPDs, $\mathcal{H}_g(x,b_{\perp})$, $\mathcal{E}_g(x,b_{\perp})$ and $\widetilde{\mathcal{H}}_g(x,b_{\perp})$ in the  impact parameter space as functions of $x$ and the transverse impact parameter $b_{\perp}$ at the scale $\mu_0^2=0.23 \sim 0.25~\mathrm{GeV}^2$. The FT is with respect to the transverse momentum transfer at zero skewness.}
		\label{gluon GPD in coordinate space}
	\end{center}
\end{figure*}

The sum rule in Eq.~\eqref{jirule}, which pertains to the forward limit of the GPDs, relates to the z-component of the total angular momentum of partons within a nucleon polarized in the z-direction. In Fig.~\ref{density Jz in valence quarks and dynamical gluon}, we plot the distribution $\mathcal{J}^z(x) = \frac{1}{2} x \left[H(x,0,0)+E(x,0,0)\right]$ as a function of $x$, displaying results for both valence quarks and gluons. The $u$ quark and the gluon contributions are positive, while the $d$ quark contribution is predominantly negative across the $x$ range. We present the $J^z$ results for valence quarks and gluons in Table~\ref{table_TAM}. The $u$ quark exhibits a positive and dominant contribution, as shown in Fig.~\ref{density Jz in valence quarks and dynamical gluon}. The $d$ quark provides a negative contribution, and the gluon contributes positively with a value of $J^z_g \approx 0.066$ at the scale $\mu_0^2=0.23\sim 0.25$~GeV$^2$, accounting for approximately 13\% of the total $J^z=0.5$. Note that at the scale $4$ GeV$^2$, the gluonic contribution to the proton total angular momentum:  $J^z_g \approx 0.194~(38\%)$~\cite{Tandy:2023zio} and $J^z_g \approx 0.187~(37\%)$~\cite{Alexandrou:2020sml} have been reported by  the DSE approach and the lattice QCD simulation, respectively.
 
\begin{table}[ht]
	\centering
	\caption{The partonic contributions to the total angular momentum of the proton.}
	\vspace{0.5em}
	\label{table_TAM}
	\begin{tabular}{c|c||c|c||c|c}
		\hline
		\hline
		$J^z_d$ & $-0.039$ & $J^z_u$ & $0.473$ & $J^z_g$ & $0.066$ \\
		\hline
		$d\%$ & $-7.8\%$ & $u\%$ & $94.6\%$ & $g\%$ & $13.2\%$ \\
		\hline
		\hline
	\end{tabular}
\end{table}

\subsection{The gluon GPDs in the impact parameter space\label{Sec4.2}}

The GPDs in the impact parameter space (IPS) provide insight into the distribution of partons with a specific longitudinal momentum fraction $x$ within the transverse position or IPS variable $b_{\perp}$. Unlike GPDs, these distributions in the IPS obey certain positivity conditions and can be given a probabilistic interpretation~\cite{Burkardt:2002hr}.

In Fig.~\ref{gluon GPD in coordinate space}, we present our results for the GPDs in the IPS, displaying them as functions of $x$ and $b_{\perp}$. We observe that the GPD $\mathcal{H}_g(x,b_{\perp})$ satisfies positivity constraints, such that $\mathcal{H}_g(x,b_{\perp}) \ge 0$, thereby permitting a probabilistic interpretation~\cite{Burkardt:2002hr}. The peak at $b_{\perp} = 0$ indicates the highest probability of finding a gluon with a momentum fraction of $x=5/16.5$. As anticipated, the gluon density gradually decreases as we move away from the proton's center ($b_{\perp} = 0$). The GPD $\mathcal{E}_g(x,b_{\perp})$ is negative, displaying a negative peak for $x=5/16.5$ at $b_{\perp} = 0$. For a probabilistic interpretation, it is essential to consider amplitudes where the initial and final states share the same helicity. However, since the GPD $E_g(x,t)$ in momentum space is associated with states possessing different helicities in the initial and final states (see Eq.~\eqref{gpd_eq2}), so that developing a probabilistic interpretation for $\mathcal{E}_g(x,t)$ is challenging. Nevertheless, a density interpretation of $\mathcal{E}_g(x,t)$ is possible by considering the superposition of transversely localized nucleon states with opposite helicities~\cite{Burkardt:2002hr}. The GPD $\widetilde{\mathcal{H}}_g(x,b_{\perp})$ in Fig.~\ref{gluon GPD in coordinate space} also obeys the positivity constraint. It represents the density difference between positive-helicity and negative-helicity gluons. For $\widetilde{\mathcal{H}}_g(x,b_{\perp})$, we find that the peak at $b_{\perp} = 0$ resides at $x=6/16.5$.

We further notice in Fig.~\ref{gluon GPD in coordinate space} that the width of all the GPDs in the transverse IPS decrease with increasing $x$. This implies that the distributions are more concentrated and the gluon is more localized near the center of momentum ($b_\perp=0$) when it is carrying a higher longitudinal momentum. Meanwhile, the peaks of all the IPS GPDs move toward to lower values of $x$ as $b_\perp$ increases. This characteristic of the GPDs in the $b_\perp$-space is reassuring since the gluon GPDs in momentum space become wider in $-t$ as $x$ increases, as can be seen from Fig.~\ref{gluon GPD in momentum space}. On the light-front, we can understand this as the larger $x$, the smaller the kinetic energy carried by the gluon. As the total kinetic energy remains limited, the distribution in the transverse momentum broadens at higher longitudinal momentum fraction reflecting the trend to  carry a larger portion of the kinetic energy.  As a consequence, these general features should be nearly model-independent characteristics of the GPDs and, indeed, they are also noticed in other theoretical investigations of the GPDs~\cite{Burkardt:2002hr,Chakrabarti:2013gra,Vega:2010ns,Mondal:2015uha,Chakrabarti:2015ama,Mondal:2017wbf,Maji:2017ill}.

\begin{figure}[!ht]
	\begin{center}
		\includegraphics[width=0.98\linewidth]{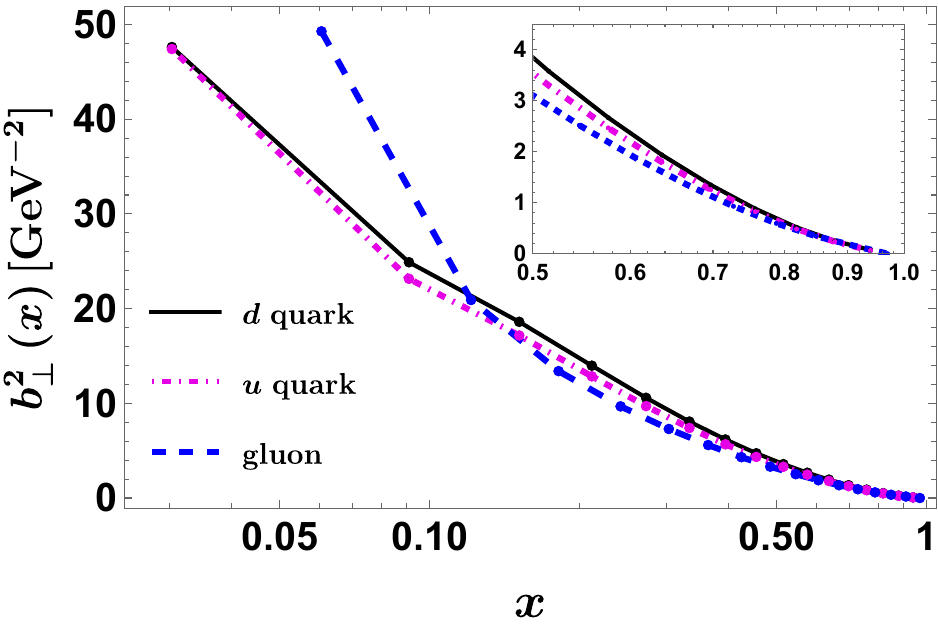}
		\caption{Plot for the $x$-dependent squared radius $b_\perp^2\left(x\right)$ of the parton density in the transverse plane. The black solid curve is for the $d$-quark, the dot-dashed magenta curve is for the $u$-quark and the blue dashed curve is for the gluon.}
		\label{squared radius dug}
	\end{center}
\end{figure}

In Fig.~\ref{squared radius dug}, we illustrate the $x$-dependent squared radius of quark and gluon densities in the transverse plane as a function of $x$. The term $\left\langle b_{\perp}^2 \right\rangle^{q/g} \left(x\right)$ characterizes the transverse size of the hadron and demonstrates an increase in the transverse radius as the parton momentum fraction $x$ decreases~\cite{Dupre:2016mai}. For a given value of $x$, the transverse size of the $u$ quark is slightly smaller than that of the $d$ quark. At lower $x$ values, the transverse size of the gluon is larger than that of the quark. However, at higher $x$ values, the gluon's transverse size is smaller than the quark's. 

According to the $x$-dependent squared radius of the proton distributions, where we compare the BLFQ prediction with available extracted data from the DVCS process within the range $0.05 \leq x \leq 0.2$~\cite{Dupre:2016mai}. The $\braket{b_\perp^2}(x)$ describes the transverse size of the nucleon and shows a decreasing value of the up quarks and down quark with increasing value of the quark momentum fraction $x$. We evaluate the proton's transverse squared radius combining the PDFs $f^q(x)$ following the reference~\cite{Dupre:2016mai}.
\begin{equation}
	\begin{split}
		\braket{b_\perp^2}=\sum_q e_q \int_0^1 \dif x f^q(x) \braket{b_\perp^2}^q(x).
	\end{split}
	\label{integration of pdf dot squared radius}
\end{equation}
In our approach, we obtain the squared radius of the proton model, $\braket{b_\perp^2}=0.473\ \mathrm{fm^2}$, around $10\%$ above the experimental data~\cite{Dupre:2016mai}: $\braket{b_\perp^2}_\mathrm{exp} = 0.43 \pm 0.01\ \mathrm{fm^2}$.


\section{Summary\label{Sec5}}
In this work, we compute the leading twist chiral-even gluon GPDs of the proton utilizing the BLFQ framework. This is achieved by numerically resolving the light-front bound state eigenvalue equation, using an effective QCD Hamiltonian incorporating three dimensional confinement in the leading Fock sector and fundamental QCD interactions for the one dynamical gluon Fock sector.

We analyze the three non-zero chiral-even GPDs at zero skewness, both in momentum space and impact parameter space. Our results show that $H_g$ and $\widetilde{H}_g$ are positive, while $E_g$ is negative. The peak magnitudes of $E_g$ and $\widetilde{H}_g$ are similar, whereas $H_g$ exhibits a notably larger peak magnitude. Comparable observations are made in the impact parameter space, where $\mathcal{H}_g$ and $\widetilde{\mathcal{H}}_g$ satisfy positivity constraints, thus offering a density interpretation.

We perform a comparison study of the transverse square radius, $\left\langle b_{\perp}^2 \right\rangle^{q/g} \left(x\right)$, for both the gluon and quarks, as a function of the longitudinal momentum fraction $x$.

As anticipated, the transverse size decreases with increasing $x$; as $x \rightarrow 1$, the transverse size of the proton behaves like a point-like object.  Additionally, using the nucleon spin sum rule, we calculate the gluon and quarks contribution to the proton's total angular momentum. The $u$ quark provides a dominant contribution, while the gluon's contribution is $J_g = 0.066$. The investigation of nonzero skewness GPDs as well as the chiral-odd sector will be the subject of future research.

\section*{Acknowledgements\label{Sec6}}
We thank Zhimin Zhu, Ziqi Zhang and Yiping Liu for helpful discussions in the Institute of Modern Physics, University of Chinese Academy of Science. S. N and C. M. thank the Chinese Academy of Sciences Presidents International Fellowship Initiative for their support via Grants No. 2021PM0021 and 2021PM0023, respectively. C. M. is supported by new faculty start-up funding from the Institute of Modern Physics, Chinese Academy of Sciences, Grant No. E129952YR0. X. Z. is supported by new faculty startup funding by the Institute of Modern Physics, Chinese Academy of Sciences, by Key Research Program of Frontier Sciences, Chinese Academy of Sciences, Grant No. ZDBS-LY-7020, by the Natural Science Foundation of Gansu Province, China, Grant No. 20JR10RA067, by the Foundation for Key Talents of Gansu Province, by the Central Funds Guiding the Local Science and Technology Development of Gansu Province, Grant No. 22ZY1QA006, by Gansu International Collaboration and Talents Recruitment Base of Particle Physics (2023-2027), by International Partnership Program of the Chinese Academy of Sciences, Grant No. 016GJHZ2022103FN and by the Strategic Priority Research Program of the Chinese Academy of Sciences, Grant No. XDB34000000. J. P. V. is supported by the Department of Energy under Grants No. DE-SC0023692 and DE-SC0023707. A major portion of the computational resources were also provided by Sugon Advanced Computing Center.


\clearpage
\biboptions{sort&compress}
\bibliographystyle{elsarticle-num}
\bibliography{ProtonGPDletter.bib}
\end{document}

%% file: preamble_math_physics.tex
\usepackage{amsmath,amssymb,mathtools} 
\ifpdftex
    \usepackage{bbm}

\else
    \usepackage[bold-style=ISO]{unicode-math}
\fi

\usepackage{siunitx}  
\makeatletter
\newcommand{\vast}[1]{\bBigg@{#1}}
\makeatother

\newcommand{\intd}[3][]{\ifthenelse{\isempty{#3}}{\mathrm{d}^{#1} #2}{\frac{\mathrm{d}^{#1} #2}{#3}}\;}

\usepackage{stackengine}

\newcommand{\cond}{\; | \;}
\newcommand{\setcond}[2]{\ifthenelse{\isempty{#2}}{\{#1\}}{\{#1\cond{}#2\}}}

\usepackage{braket}
\usepackage{cancel}
    
